\documentclass[sigconf,nonacm]{acmart}

\usepackage{tikz}

\AtBeginDocument{%
  \providecommand\BibTeX{{%
    \normalfont B\kern-0.5em{\scshape i\kern-0.25em b}\kern-0.8em\TeX}}}

\makeatletter
\def\@ACM@checkaffil{
    \if@ACM@instpresent\else
    \ClassWarningNoLine{\@classname}{No institution present for an affiliation}%
    \fi
    \if@ACM@citypresent\else
    \ClassWarningNoLine{\@classname}{No city present for an affiliation}%
    \fi
    \if@ACM@countrypresent\else
        \ClassWarningNoLine{\@classname}{No country present for an affiliation}%
    \fi
}
\makeatother

\begin{document}

\title{SICP Industry Summit 2026-07: \\ AI and Secure Software Development Summit}

\author{Anna Lena Rotthaler$^{\dagger\ddagger}$, Yasemin Acar$^{\dagger\ddagger}$, Stefanie Borchert$^{*}$, Simon Oberthür$^{\dagger\ddagger}$, Elisabeth Ostryanin$^{\dagger}$, Dana Rohmann$^{\dagger}$, Juraj Somorovsky$^{\dagger\ddagger}$, Simon Trang$^{\dagger\ddagger}$}

\affiliation{
    \institution{$^{\ddagger}$Software Innovation Lab , Paderborn University, Paderborn, Germany}    
}

\affiliation{
    \institution{$^{*}$Leibniz University Hannover, Germany}    
}

\affiliation{
    \institution{$^{\dagger}$Paderborn University, Paderborn, Germany}
}

\renewcommand{\shortauthors}{Paderborn University | SICP}
\renewcommand{\shorttitle}{S3C2 and SICP Summit 2026-07: AI and Secure Software Development Summit}

\begin{abstract}
  AI capabilities have widely increased and sped up cyber attacks targeting less secure elements in the software supply chain and causing significant damage to businesses and organizations. AI as part of the supply chain also increases products' attack surface. On July 8, 2026, researchers from Paderborn University, also in the context of SICP\,\textemdash\,Software Innovation Campus Paderborn conducted an AI and Secure Software Development Summit with a diverse set of 12 practitioners from 12 companies, modelled after S3C2 Secure Software Supply Chain summits. The goal of the Summit is to enable sharing between industry practitioners having practical experiences and challenges with AI and secure software development, and helping to form new collaborations.
  We conducted six panel discussions based on open-ended questions regarding 
  experiences with code generation, supply chain security, quality improvement of existing code by AI, AI as part of products, AI as an external threat, and agentic AI security.
  The open discussions enabled mutual sharing and shed light on common challenges that industry practitioners with practical experience face when dealing with AI in secure software development. In this paper, we provide a summary of the Summit. Full panel questions can be found in the appendix.
\end{abstract}


\maketitle

\begin{tikzpicture}[overlay, remember picture]
\node[anchor=north west, 
      xshift=15.5cm, 
      yshift=-2.1cm] 
     at (current page.north west) 
     {\includegraphics[width=4.1cm]{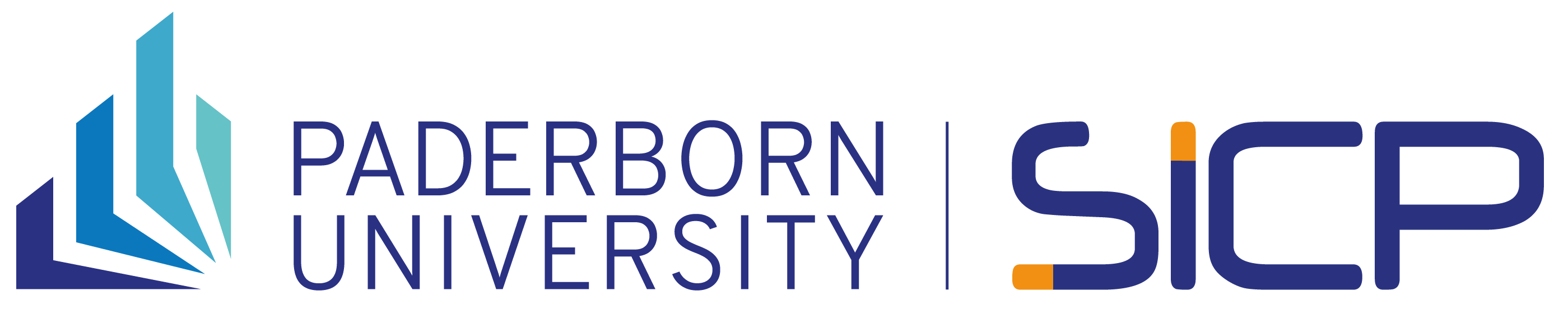}}; 
\end{tikzpicture}

\section*{Introduction}

  AI capabilities have widely increased and sped up cyber attacks targeting less secure elements in the software supply chain and causing significant damage to businesses and organizations. AI as part of the supply chain also increases products' attack surface. On July 8, 2026, researchers from Paderborn University, also in the context of SICP\,\textemdash\,Software Innovation Campus Paderborn conducted an AI and Secure Software Development Summit with a diverse set of 12 practitioners from 12 companies, modelled after S3C2\footnote{\url{https://s3c2.org}} Secure Software Supply Chain summits~\cite{rahman2026s3c2, rotthaler2025s3c2, enck2022top}. The goal of the Summit is to enable sharing between industry practitioners having practical experiences and challenges with AI in secure software development; to help form new collaborations between industrial organizations and researchers; and to identify research opportunities.

  Summit participants were recruited from 12 companies, intentionally in diverse domains and having various company maturity levels and sizes.  Attendance is limited to one participant per company to keep the event small enough that honest communication between participants can flow. The Summit was conducted under the Chatham House Rules, which state that all participants are free to use the information discussed, but neither the identity nor the affiliation of the speaker(s), nor any other participant, may be revealed. As such, none of the participating companies are identified in this paper.  

  The Summit consisted of one keynote presentation and six panels. Before the Summit, participants completed a survey to vote on the topics of the six panels.  As such, the panel topics represent the challenges faced by practitioners. Based upon personal preferences expressed in the survey, three to four participants were selected to begin each 45-minute panel discussion with a 3-5 minute statement.  The remaining minutes of each panel were spent openly discussing the topic. The questions posed to the panelists appear in the Appendix.   

 Researchers (three professors, one PhD student, two Master's student and two research associates) took notes on the discussions. We jointly created a first draft summary of the discussion based on these notes. The draft was reviewed by all attendees and authors.

The next six sections provide a summary of the AI and Secure Software Development Summit.

\section{Code Generation}
Attendees reported that ``everyone uses AI for code generation'' and discussed that Ai enabled work on projects that may otherwise not get done. The process of integrating AI use varies between companies. Processes included being told to ``familiarize yourself with it'', building a team-based skill base, and being asked to ``start using AI next week''. One attendee explained how they classify four layers of AI-integration: 1) use ChatGPT (or any chat-based LLM) and ask questions (which they said can work better than Google search), 2) use Copilot in the IDE to ask questions in the IDE and for auto completion, 3) use AI in the IDE with an agent that has access to files or projects (guided programming, in this case using agentic AI), and 4) full agentic (team of agents gets specifications and tasks).
Attendees expressed a wish for within-company instruction and guidance for AI usage.

\subsection{Quality \& Risks}
Attendees reported that more important than the choice of the model are the skills provided to the model (i.e., instruction bundles).
They stated that the integration layer determines how effective AI use can be; a deeper layer means more trust is needed.
Attendees mentioned a quality risk: if less experienced developers generate code that they cannot themselves meaningfully review, they burden experts with the large volume of generated code.
Attendees emphasized that expert review and correction remain necessary when using AI-generated code.

Attendees discussed that AI usage for code generation creates the need for experts to fix the AI-generated code and thus the need for experts.
They reported that AI-generated output cannot be trusted without checks; an example mentioned was that AI slop filled with emojis appeared in unchecked technical documentation.

Attendees reported that AI introduces new security risks and hallucinations. An attendee said that ``you must be willing to take risks if you want to get the best out of AI''.

AI was mentioned as an important tool for generating tests that would not have been written without AI. At the same time, attendees reported that while AI can review code and sometimes find bugs, they may have also found the same bugs in static code analysis.

\subsection{Costs}
When talking about AI and costs, the term ``wild west'' was mentioned multiple times.
Oftentimes, it is unclear how much cost is caused by AI.
Attendees reported that the expectation when using AI is often ``AI means better, faster, cheaper,'' but that this is not always true.
It was mentioned that AI cost increased from July 1 2026, and could go up again.
Attendees explained that AI currently does increase speed but does not decrease cost, but they expect it to be a game changer in the future.
An estimate cited during the discussion suggested a 40\% efficiency improvement alongside a 60–70\% increase in risk-related costs associated with hallucinations. Additional development time needed to evaluate and correct AI-generated code was discussed as a contributing factor.
Attendees agreed that cost does not increase because of code generation but because of validation and certification, and that validation is more important than generation.
They reported different ways of handling costs: from a certain amount of tokens to as many new tokens as needed. One attendee explained that if their tokens are empty, they use free local LLMs.
Another attendee highlighted the importance of tokens by saying: ``If tokens are empty, it causes similar trouble as if the coffee machine is not working.''
Attendees agreed that token consumption should be transparent.
The discussions also highlighted approaches to structuring and validating AI-assisted development. Suggestions included breaking work into small, well-specified tasks, defining explicit acceptance criteria, and keeping human-defined tests unchanged when evaluating AI-generated code. Participants raised the concern that AI-generated code could satisfy AI-generated tests without meeting the intended requirements. The extent of validation and human-in-the-loop involvement should reflect the criticality of the software, with deterministic checks and explicit quality gates discussed as safeguards for critical applications.

\subsection{Open Questions}
At the end of the panel, some open questions and problems remained:

\begin{itemize}
    \item How do you effectively organize ``change management''?
    \item How do you convince teams to use AI, and what do you do with people who don’t want to ``go along with it''?
    \item How do you get a handle on follow-on costs and risks (as a trade-off for faster development)? Cost shifting and managing it. Communicate this to external customers as well?
    \item Generation is going well—but the effort required for validation and testing remains high. How can this be improved?
    \item What’s the best way to integrate junior developers into development and company processes? (Due to the use of AI and rapid developments, it’s extremely difficult for them to get started.)
    \item How much more effective/efficient are employees with AI?
\end{itemize}

\section{Supply Chain Security}
In this panel, attendees discussed how their supply chains changed because of AI, which components they trust, and why and how responsibilities and internal processes changed.

Some attendees reported that shadow IT makes it difficult to maintain visibility into their software supply chains, including dependencies, licenses, and vulnerabilities.

Others reported that past incidents (e.g., Log4j) had catalyzed centralizing information about the supply chain. 

In a discussion about trust in their components, some attendees said that they do not use the latest version of a component immediately after release and wait for vulnerabilities to be found.

Talking about how plugins get qualified to be used in projects, one attendee said that some amount of trust is necessary because reviewing everything takes unreasonably long.

Attendees talked about how to manage the wave of vulnerabilities.
One attendee reported considering automating the evaluation of issues to handle the huge number of vulnerabilities.
That means an AI categorizes the issues beforehand, and a human can decide which issues need to be fixed and how.
It is possible that their code evades the vulnerability in their dependency.
Others mentioned to reconsider which dependencies they can do without and reduce the number of vulnerabilities that way.

For vulnerabilities in the Linux kernel, one attendee said that the kernel has too many vulnerabilities to patch them all; you can only learn to accept some of them.

Attendees discussed whether and how NIS2 and DORA get checked.
Strategies ranged from doing audits to having SBOMs and quality gates checked by external parties.
One attendee added that this depends on their customers.

\subsection{Open Questions}
At the end of the panel, some open questions remained:

\begin{itemize}
    \item How do AI and dependencies (including hallucinated ones that can be squatted) work?
    \item What predefined context safeguards the supply chain?
    \item How do we deal with the massive wave of security alerts (CVEs)? Is the current system still adequate?
    \item How can we effectively prioritize and assess the many new vulnerabilities automatically?
    \item How do regulations such as NIS2, CRA, DORA, and the AI Act affect supply chain security?
\end{itemize}

\section{Quality Improvement of Existing Code with AI}
In this panel, attendees discussed how they use AI to improve existing code.
They reported that AI helps with code analysis when the size of projects is too large to handle for employees.
It can do a broad analysis of the code and create lists for human coders to shift their focus towards the right problems.
Other attendees reported that they use AI to help migrate their code.

Some attendees argued that AI reduces human bias, which means that humans might implicitly trust certain sections of code while an AI treats each piece of code equally.
Other attendees argued that based on your prompt and personal context with the AI, it might add other kinds of bias into the code analysis, sometimes even subconsciously.
They reported that depending on how you ask an AI might find solutions to problems that do not have a solution.

One attendee said that AI suggests improvements in code that has worked for years without any issues. They argued that some things do not need to get fixed just because an AI said that there is an issue. 
They said you should not waste resources on issues that no one complained about.
Another attendee recounted an incident where they were looking for the root of an issue, and AI was able to find the cause in a piece of code that was old and implicitly trusted.

One attendee raised the question whether code has to be human-readable if it gets reviewed by AI anyway.
Another attendee reported that common good practices like ``Keep it Simple'' are not just good for human reviewing but also help the AI.

\subsection{Change Management}
In a discussion about how to get your employees to use AI efficiently, one attendee proposed to write down knowledge to share it with AI.
Furthermore, they agreed that you need to find leaders in your company who are AI experts---possibly self-taught---and organize for them to teach and motivate others, which can be difficult because experts may be more interested in advancing their own work instead of educating others.
Another attendee stressed that intrinsic motivation is important. 
Attendees reported that some people deeply reject AI in their workflow, for example, for environmental reasons or because they simply enjoy the process of programming.
Those employees need to be assigned to tasks that ask for a human, which exist still.
But the demand for traditional coding is much smaller than before.

They agreed that some people need some time to become comfortable using AI. 
One attendee added that some people are not aware of what AI is capable of, some do not even know how to prompt effectively.
They might need a demonstration to see what an AI can do if prompted correctly.

Attendees reported that in the beginning, their workflow was slowed down because they did not know how to use AI, but after some time, they were a lot more productive, but some time after that, they feel like they plateau,
and they might drop using AI because they feel like AI is not improving their workflows.
One attendee wondered how to capture the feeling of improving AI use so that users are inclined to stick with AI and build a habit.

One attendee added that change management needs to be done in three steps: First, management has to agree and stand wholeheartedly behind this effort, especially with financial support, for example when tokens run out, to reduce frustration. Secondly, convince leaders within the company to adopt AI. Lastly, use the leaders as multipliers in your company to teach their colleagues how to use AI. 
If those leaders are convinced to adopt AI, they can teach and convince many more people in a short time.

\subsection{Open Questions}
At the end of the panel, some open questions remained:

\begin{itemize}
    \item It’s primarily a social problem: ``How do we get people to… ?''
    \item Does code still need to be human-readable if everything is going to be read by AI anyway?
    \item How do we effectively pass on knowledge and skills?
    \item How do we handle AI in brownfield projects? Analysis is expensive (but may have already been tested); regenerating code is inexpensive.
    \item How can program code be analyzed in a modular, efficient, and effective way?
    \item How can we protect ourselves from AI gaslighting?
    \item How do we deal with documentation debt and chaotic backlogs? (Make them modular but how can we do that efficiently?)
\end{itemize}

\section{AI as part of products}
In this panel, attendees discussed AI as part of their products and how they secure it against external threats and prompt injections.
Some of the attendees do not offer LLMs as part of their products since it is not applicable to their industry, but those that do have policies in place that make sure that user data is anonymized or that there is no unauthorized access to data.

Attendees discussed how they monitor which data is put into their LLMs. They reported that they restrict who can add new data with access groups and authorization mechanisms.

Talking about how attendees verify that their AIs are safe from prompt injection, one attendee reported that they use logging of input and output, and another AI checks and summarizes those logs to find suspicious behavior.
Those logs are also anonymized to ensure the privacy of the users and employees.
Some attendees said that they use another AI in the middle that decides if a task can be solved by a cheaper model or if it is complex enough that it requires an expensive model to minimize their cost in regards to how many tokens are used.

Attendees discussed how they feel pressure from their customers to implement AI in their products but that it is difficult to realize those demands
because AI comes with a lot of responsibility to protect their users.
There are many security threats involved with AI; one attendee suggested using deterministic checks to combat these issues.

\subsection{Open Questions}
At the end of the panel, some open questions remained:

\begin{itemize}
    \item How can you minimize the potential negative impacts of AI in your own products?
    \item How can data security and anonymization be ensured?
    \item How can token costs be controlled?
    \item How should pre-checks be effectively designed? (e.g., safeguards)
    \item How should monitoring be set up?
    \item How should access controls be defined and implemented?
\end{itemize}

\section{AI as an external threat}
In this panel, attendees discussed how AI facilitates attacks on existing systems. In the examples discussed, AI primarily increased the speed and complexity of attacks using established attack vectors, reducing the time to exploit.

\subsection{Experiences with AI Attacks}
Attendees agreed that AI only dramatizes the actual problem: comprehensive component updates are not possible because everything needs to be double checked and that is not possible with the required speed. Updates must be validated and arranged with the customer, thus, the frequency of updates depends on the risk. One attendee highlighted that daily updates are not possible for every system and customer.

Attendees concluded that the frequency of updates needs to be increased but stated that the supply chain is not that fast.
One attendee reported that they wait a couple of days before updating, but then they are vulnerable for multiple days because there are only hours between CVE and exploit. The solution to this problem was unclear; attendees suggested that there is no perfect solution.

AI was called a powerful weapon for attackers: potential attackers need little research and understanding to use AI for attacks. Social engineering attacks are facilitated by AI as they can easily be conducted in multiple languages. Attendees discussed that bots can make ``realistic'' phone calls and simulate humans (e.g., keyboard noise).

One attendee reported that the attack surface is smaller for backends.

\subsection{Dealing with Vulnerabilities}
Attendees discussed how they deal with vulnerabilities and how difficult it is to apply security updates. They reported that updates are very slow in operational technology (OT) because they cause production downtime and thus high costs.
One attendee mentioned that there is no possibility for quick updates for embedded systems, because whole sections must be changed instead of individual files.
The complexity of updates differs by product, some products need to be updated on-site.
Attendees also reported technical limits of updates, e.g., they can be limited by storage capacities, which leads to treating the same bug differently on different hardware.

\subsection{Open Questions}
At the end of the panel, some open questions remained:

\begin{itemize}
    \item Time-to-exploit is getting shorter and shorter (on average less than an hour). Time-to-patch cannot be reduced to that time frame. How can the risk of exploits be minimized? (Risk management)
    \item How can OT Security keep up?
\end{itemize}

\section{Agentic AI Security}
All attendees said their companies have AI usage policies. However, attendees reported that these policies are lagging behind actual practices in development and the fast-changing landscape of new AI tools. Attendees reported that these policies are important for oversight of data security, especially for IP and confidentiality protection when using new tools.

Attendees agreed that AI agents should be contained through guardrails like sandboxing with access control via MCP (model context protocol, open standard by Anthrophic), dedicated servers, and containers without internet access, testing agents only on non-critical systems. 
Developers should contemplate the extent of agents and human-in-the-loop. Attendees reported decision fatigue when monitoring AI, leading to default trust in AI outputs. Agents surveilling agents was a recommended safeguarding system that multiple attendees envisioned, with a third controlling entity.

Multiple attendees reported the risk of potential token over-consumption by agents, which has to be contained. One attendee presented two orchestration patterns:

\textbf{Internally steered}: Two implementations for the same project with different loops, carried out by individual agents that are internally managed from one orchestrator.

\textbf{Externally steered}: Orchestration by one agent for specifications and parallelization of agents, without sub-agents, all steered from outside without trust in external sources.

The attendee remarked that AI vendors tend to manage everything in a black-box and in-house orchestrators fall behind. This makes standardizing AI harnesses costly, but worth it.
One company already uses AI agents for improving security; the attendee stressed the need for security standards and a human-in-the loop to monitor agents' actions.

\subsection{Open Questions}
At the end of the panel, some open questions remained:

\begin{itemize}
    \item How can we set reasonable limits (sandboxing, human-in-the-loop, guardrails), balancing delays and security?
    \item Data protection?
    \item How can we achieve effective regulation (for security) without unnecessary delays?
    \item How can regulations be developed quickly enough so that they don’t lag behind technology and don’t lead to shadow IT?
    \item How can we conduct effective and dynamic threat modeling and use it to create appropriate regulations?
\end{itemize}

\section{Executive Summary}
\label{executive}
AI is used---among others---by software developers as well as by attackers. The time-to-exploit is getting shorter and shorter but the time-to-patch cannot keep up.
Attendees discussed their usage of AI for code generation and testing.
Open problems include how to continue to train junior roles in the workforce and onboard senior roles to current work practices, how to keep up with reviews and tests of generated code, and how to maintain and increase product and organizational security.

\begin{acks}
A big thank you to all Summit participants. We are very grateful for being able to hear about your experiences and suggestions. The Summit was organized at Software Innovation Campus Paderborn by Yasemin Acar, Simon Oberthür, Juraj Somorovsky, Simon Trang, and Anna Lena Rotthaler, modeled on previous summits organized by Laurie Williams, and was recorded by Dana Rohmann, Elisabeth Ostryanin, Anna Lena Rotthaler, Juraj Somorovsky, Simon Oberthür, Stefanie Borchert, and Yasemin Acar. The summit was sponsored by Paderborn University in the context of SICP. This material is based upon work supported by the National Science Foundation Grant Nos. 2207008, 2206859, 2206865, and 2206921.
These grants support the Secure Software Supply Chain Summit (S3C2), consisting of researchers at North Carolina State University, Carnegie Mellon University, University of Maryland, George Washington University, and Paderborn University. Any opinions expressed in this material are those of the author(s) and do not necessarily reflect the views of the National Science Foundation nor Paderborn University.
\end{acks}

\bibliographystyle{ACM-Reference-Format}
\bibliography{literature}

\appendix

\section{Full Survey Questions for Panel}
\label{questions}
\begin{enumerate}
\item Using AI to generate code. What is your experience with quality and effort? Security, data protection? How do you test code developed with AI support? Do you use AI to support testing? Do you generate tests with AI as part of the development process? How well does that work for you? Costs? Process security? Human-in-the-loop? Does usability change, for users, for agents?

\item How has your supply chain changed? How can you trust your suppliers? Responsibilities, safeguards, internal processes? Have you established any cooling-off periods?

\item What are your experiences using AI to find vulnerabilities (and possibly prioritize and patch them)? Does AI help you make sense of the results from analysis tools? How do you test code that was developed with AI support? Do you use AI to assist with testing? Do you use AI to generate tests? How well does that work for you? How effective is AI in helping to solve problems in code? 

\item How do you protect AI-based applications you offer as products against external threats? Do you take measures to protect against prompt injection?

\item Agentic AI can be used to attack systems. Do you have any experience with this, including incident reporting, defense, and prevention? What types of security vulnerabilities are taken into account? (How) do you use network monitoring, email analysis? 

\item Secure integration of agents into company processes. Agents can perform multi-step tasks, use APIs and external tools, collaborate with other agents. How do you integrate this into your operations? What are your experiences, successes, or challenges? What security vulnerabilities are taken into account? What protective measures do you use? How are AI agents’ actions logged/monitored? How much human-in-the-loop is appropriate? Discussion may be broader than software development, about business processes in general. What about governance? 
\end{enumerate}

\end{document}